\documentstyle[11pt]{article}
\begin{document}
\title{The ambiguous meaning of irreversibility}
\author{X. de Hemptinne\\{\it Department of
Chemistry, Catholic University of Leuven,}\\
{\it Celestijnenlaan 200 F, B-3001 Heverlee, Belgium}}
\maketitle

\medskip

\begin{abstract}
Irreversibility of spontaneous macroscopic dynamics and its
asymmetry with respect to the sign reversal of the variable $t$ is
usually interpreted as a genuine property of complex isolated
systems. Discussion of the kinetics involved in Joule's experiment
concerning spontaneous expansion of a gas shows that the isolation
hypothesis results from ambiguous definitions of a number of
keywords. Whereas Poincar\'{e}'s apparent irreversibility result
from conservative Hamiltonian dynamics, full relaxation implies
interaction with the outside world. Only the latter process leads
to entropy change.
\end{abstract}

\medskip
\centerline{\bf 1.~Introduction}
\medskip

Relating conservative Hamiltonian mechanics valid at the
microscopic level to the memory loosing property of relaxations
characterizing macroscopic systems remains a challenging exercise.
Ambiguous definitions turns this matter into a controversial
debate.

Irreversibility is traditionally illustrated by considering a
double box, half filled with gas and half evacuated, with a
membrane separating the two parts [1]. This is a simplified
implementation of Joule's celebrated experiment where he
demonstrated that spontaneous expansion of an ideal gas in vacuum
from one equilibrium state to a new one occurred without net
exchange of heat with the neighbourhood. Comparing the conditions
at the start and at the end of the expansion, which are clearly
two equilibrium states, but without entering into the detailed
mechanism involved, he concluded that there had been no
interaction with the environment. Since then it has been assumed
without further discussion that spontaneous relaxation from one
equilibrium condition to a new one is a genuine property of
isolated systems, the latter term implying strict conservation of
energy (elastic collisions with the walls) and matter. In order to
evaluate the pertinence of this conclusion, an experiment
concerning the dynamics involved is necessary.

\medskip
\centerline{\bf 2.~Joule's experiment}
\medskip

Firing a compressed air pistol in a room is, among others, an
adequate experiment to test the dynamics of spontaneous expansion.
This may be done either in an acoustic reverberation room or in an
anechoic chamber.

In the first case, brutal expansion of the air generates a violent
and long lasting acoustic perturbation. In the second case, this
is almost absent. The pistol and the air are the same and so are
their mechanical properties. The only difference between the two
experiments is the nature of the walls. The experiments show
therefore unambiguously that the global dynamics leading to final
equilibrium implies somewhere interaction with the walls.

In Joule's conclusion, the word ``equilibrium'' clearly points to
conditions reached when the system has been allowed to relax. By
this he assumes conditions where all collective motion has died
out, including acoustic perturbations.

\medskip
\centerline{\bf 3.~Mechanism}
\medskip

The obvious outcome of the experiment is that the global mechanism
of spontaneous expansion of a gas requires at least two steps, one
of which implying interaction with the neighbourhood. The kinetics
is dominated by the slowest or rate determining step, which is
different in the two cases mentioned above.

Depending on the mechanical properties of the walls, the two steps
may be almost concomitant or well separated in time. For
simplicity, in the discussion to follow it will be assumed that
they are separated.

As soon as the constraint defining the initial conditions is
removed, a jet is created that turns soon into an acoustic
perturbation by reverberation on the walls. The jet is a
collective (alias: coherent) motion. Energy borrowed from the
thermal supply available at the onset is transferred partially
into this motion.

The acoustic perturbation growing by alteration of the jet remains
non-thermal, although implying possibly extremely complex
molecular trajectories. This modified perturbation may therefore
still be given the predicate collective (or coherent). The
transformation is conservative (implies elastic collisions at the
walls).

Although the particles of the gas disseminate throughout the
volume in a motion that may be chaotic, the memory of the initial
conditions is preserved. The amplitude and the phase relation
between the components of the acoustic spectrum are its signature.

The final step of the global process concerns destruction of the
motion's coherence by collisional exchange of fluctuations with
the surrounding thermal bath. Friction and other surface forces
causing acoustic absorption by the boundaries belong to the same
mechanism. By resolving the correlation present in the collective
motion, the energy that had been diverted initially is restored
progressively in the system's incoherent thermal bath. At the end
of the process, when the new state of equilibrium has been
reached, the energy in the thermal bath is as before, giving the
illusion that there had been no exchange with the surroundings.

\medskip
\centerline{\bf 4.~Irreversibility}
\medskip

Irreversibility suggests non-recurrence of initial singular
events. With macroscopic systems, due to the extremely large
Poincar\'{e} time, initial dissemination responds perfectly to
this definition. In mechanics however, the word bears also a more
subtle connotation, indicating that the relevant dynamics is
asymmetric with respect to the fictitious sign reversal of the
variable $t$. Let the definitions be labelled respectively
``weak'' and ``strong''. According to them, dissemination governed
by Hamiltonian dynamics is weakly irreversible. Being dominated by
stochastic interventions of the environment, the second step in
the global process is clearly strongly irreversible.

Dissemination of particles under Hamiltonian dynamics has been the
object of formal treatment in the recent decennia [2] in the
context of ergodic theory and mixing. This has stimulated an
abundant literature argumenting that chaos generating
perturbations justify diffusion-like properties. The logics is
however constructed on an alternative criterion for equilibrium
where the obvious presence of the collective mode is waved aside.
No matter how chaotic the motion may be, Hamiltonian dynamics
alone does not remove the strong time correlation inscribed in the
motion. An additional mechanism is required. This implies
stochastic dissipation to the environment if the system is to
reach full thermodynamic equilibrium.

\medskip
\centerline{\bf 5.~Entropy}
\medskip

Boltzmann's entropy is related to the measure of the part of phase
space available to the system. In a microcanonical context, where
isolation is assumed, entropy is an explicit function of the
collection of extensive properties defining the system's
conditions. With a gas at equilibrium, the traditional variables
are the energy $E$, the volume $V$ and the number of particles
$N$. Definition of the entropy may be generalized to non-
equilibrium conditions by including the additional constraints as
new variables.

In the first step of the relaxation process, work performed by the
system on itself by expanding $(p{\rm d}V)$ converts thermal
energy into a collective motion. The intermediate state thereby
reached is not at equilibrium. By considering energy contained in
the collective mode as the additional variable to be included in
the definition of $S$, it may be shown [4] that the entropy
function differentiates as follows:
$${\rm d}S={{{\rm d}E}\over T}+{p\over T}{\rm d}V-
\sum_k{{\mu_k}\over T}
{\rm d}N_k- {1 \over T} {\rm d}(collective\;energy).
$$
Exact balance of the work performed by the system on itself and
the energy increment in the collective mode neutralizes the
relevant terms in the latter equation. The initial change is
therefore adiabatic (${\rm d}S=0$), in agreement with Liouville's
theorem.

At the end of the first period the system is not at equilibrium.
Relaxation of the collective mode implies stochastic coherence
breaking intervention of the environment. Now the entropy
increases while the collective mode progressively vanishes.

\medskip
\centerline{\bf 5.~Conclusion}
\medskip

Relaxation implies two independent steps. The first one is
Hamiltonian. No matter how chaotic the motion may be, it is
irreversible only in Poincar\'{e}'s sense. This step is iso-
entropic. Joule's experiment shows that the obvious collective
transient generated by the initial dissemination requires
stochastic intervention of the surroundings for its own relaxation
with entropy creation. The source of strong irreversibility is
therefore external to relaxing systems.

\end{document}